\documentclass[letterpaper, amsfonts, superscriptaddress, amssymb, amsmath, preprint, showkeys, nofootinbib, raggedbottom]{revtex4-1}
\usepackage[english]{babel}
\usepackage[utf8]{inputenc}
\usepackage[colorinlistoftodos, color=green!40, prependcaption]{todonotes}
\usepackage{amsthm}
\usepackage{mathtools}
\usepackage{physics}
\usepackage{xcolor}
\usepackage{graphicx}
\usepackage{adjustbox}
\usepackage{placeins}
\usepackage[T1]{fontenc}
\usepackage{lipsum}
\usepackage{csquotes}
\usepackage{float}
\usepackage{afterpage}
\usepackage{gensymb}
\usepackage[version=4]{mhchem}

\newcommand{\tw}{_\textrm{2}}

\usepackage[pdftex, pdftitle={Article}, pdfauthor={Author}]{hyperref} 
\graphicspath{ {Figures/} }

\begin{document}
\title{Nodal fermions in a strongly spin-orbit coupled frustrated pyrochlore superconductor}

\affiliation{Department of Physics, Massachusetts Institute of Technology, Cambridge, MA 02139, USA}
\affiliation{Center for Correlated Electron Systems, Institute for Basic Science (IBS), Seoul, 08826, Korea}
\affiliation{Department of Physics and Astronomy, Seoul National University, Seoul, 08826, Korea}
\affiliation{Center for Theoretical Physics (CTP), Seoul National University, Seoul, 08826, Korea}
\affiliation{Center for Complex Phase Materials, Max Planck POSTECH/Korea Research Initiative, Pohang, Republic of Korea}
\affiliation{Advanced Light Source, Lawrence Berkeley National Laboratory, Berkeley, CA 94720, USA}
\affiliation{Faculty of Applied Physics and Mathematics, Gdansk University of Technology, Narutowicza 11/12, 80-233 Gdansk, Poland}
\affiliation{Advanced Materials Centre, Gdansk University of Technology, Narutowicza 11/12, 80-233 Gdansk, Poland}
\affiliation{These authors contributed equally}

\author{Dongjin Oh}
\email{djoh@mit.edu}
\affiliation{Department of Physics, Massachusetts Institute of Technology, Cambridge, MA 02139, USA}
\affiliation{These authors contributed equally}

\author{Junha Kang}
\affiliation{Center for Correlated Electron Systems, Institute for Basic Science (IBS), Seoul, 08826, Korea}
\affiliation{Department of Physics and Astronomy, Seoul National University, Seoul, 08826, Korea}
\affiliation{Center for Theoretical Physics (CTP), Seoul National University, Seoul, 08826, Korea}
\affiliation{These authors contributed equally}

\author{Yuting Qian}
\affiliation{Center for Correlated Electron Systems, Institute for Basic Science (IBS), Seoul, 08826, Korea}
\affiliation{Department of Physics and Astronomy, Seoul National University, Seoul, 08826, Korea}
\affiliation{Center for Theoretical Physics (CTP), Seoul National University, Seoul, 08826, Korea}

\author{Shiang Fang}
\affiliation{Department of Physics, Massachusetts Institute of Technology, Cambridge, MA 02139, USA}

\author{Mingu Kang}
\affiliation{Department of Physics, Massachusetts Institute of Technology, Cambridge, MA 02139, USA}
\affiliation{Center for Complex Phase Materials, Max Planck POSTECH/Korea Research Initiative, Pohang, Republic of Korea}

\author{Chris Jozwiak}
\affiliation{Advanced Light Source, Lawrence Berkeley National Laboratory, Berkeley, CA 94720, USA}

\author{Aaron Bostwick}
\affiliation{Advanced Light Source, Lawrence Berkeley National Laboratory, Berkeley, CA 94720, USA}

\author{Eli Rotenberg}
\affiliation{Advanced Light Source, Lawrence Berkeley National Laboratory, Berkeley, CA 94720, USA}

\author{Joseph G. Checkelsky}
\affiliation{Department of Physics, Massachusetts Institute of Technology, Cambridge, MA 02139, USA}

\author{Liang Fu}
\affiliation{Department of Physics, Massachusetts Institute of Technology, Cambridge, MA 02139, USA}

\author{Tomasz Klimczuk}
\affiliation{Faculty of Applied Physics and Mathematics, Gdansk University of Technology, Narutowicza 11/12, 80-233 Gdansk, Poland}
\affiliation{Advanced Materials Centre, Gdansk University of Technology, Narutowicza 11/12, 80-233 Gdansk, Poland}

\author{Michal J. Winiarski}
\affiliation{Faculty of Applied Physics and Mathematics, Gdansk University of Technology, Narutowicza 11/12, 80-233 Gdansk, Poland}
\affiliation{Advanced Materials Centre, Gdansk University of Technology, Narutowicza 11/12, 80-233 Gdansk, Poland}

\author{Bohm-Jung Yang}
\email{bjyang@snu.ac.kr}
\affiliation{Center for Correlated Electron Systems, Institute for Basic Science (IBS), Seoul, 08826, Korea}
\affiliation{Department of Physics and Astronomy, Seoul National University, Seoul, 08826, Korea}
\affiliation{Center for Theoretical Physics (CTP), Seoul National University, Seoul, 08826, Korea}

\author{Riccardo Comin}
\email{rcomin@mit.edu}
\affiliation{Department of Physics, Massachusetts Institute of Technology, Cambridge, MA 02139, USA}


\maketitle

\textbf{Abstract}

\textbf{The pyrochlore lattice, a three-dimensional network of corner-sharing tetrahedra, is a promising material playground for correlated topological phases arising from the interplay between spin-orbit coupling (SOC) and electron-electron interactions. Due to its geometrically frustrated lattice structure, exotic correlated states on the pyrochlore lattice have been extensively studied using various spin Hamiltonians in the localized limit. On the other hand, the topological properties of the electronic structure in the pyrochlore lattice have rarely been explored, due to the scarcity of pyrochlore materials in the itinerant paramagnetic limit. Here, we explore the topological electronic band structure of pyrochlore superconductor RbBi$_{\textrm{2}}$ using angle-resolved photoemission spectroscopy. Thanks to the strong SOC of the Bi pyrochlore network, we experimentally confirm the existence of three-dimensional (3D) massless Dirac fermions enforced by nonsymmorphic symmetry, as well as a 3D quadratic band crossing protected by cubic crystalline symmetry. Furthermore, we identify an additional 3D linear Dirac dispersion associated with band inversion protected by threefold rotation symmetry. These observations reveal the rich non-trivial band topology of itinerant pyrochlore lattice systems in the strong SOC regime. Through manipulation of electron correlations and SOC of the frustrated pyrochlore lattices, this material platform is a natural host for exotic phases of matter, including the fractionalized quantum spin Hall effect in the topological Mott insulator phase, as well as axion electrodynamics in the axion insulator phase.}

\section*{Main text}

The discovery that phases of matter can be classified by their topological invariants and crystalline symmetries is one of the greatest innovations in modern quantum material research \cite{narang_topology_2021,wieder_topological_2021}. Thanks to these methodological advances, it has become an important mission to categorize materials into various topological classes \cite{bradlyn_topological_2017,vergniory_complete_2019,zhang_catalogue_2019-1,tang_comprehensive_2019-1,xu_high-throughput_2020-1}. Over the past years, several topological states -- topological insulators \cite{fu_topological_2007-1}, topological crystalline insulators \cite{fu_topological_2011}, topological Dirac \cite{yang_classification_2014,liu_discovery_2014,liu_stable_2014,xu_observation_2015} and Weyl semimetals \cite{xu_discovery_2015,liu_magnetic_2019} -- have been extensively classified according to their inherent crystalline symmetries. Recently, the emergent non-trivial topology of quantum materials with geometrically frustrated lattices, such as massive Dirac fermions and topological flat bands in the Kagome metals \cite{ye_massive_2018,kang_dirac_2020,kang_topological_2020}, has drawn significant attention. Nonetheless, there are still numerous topological phases yet to be experimentally realized. Therefore, continuing experimental exploration of uncharted quantum topological states is of great significance toward a complete catalog of topological classification.

The pyrochlore lattice, comprising a network of corner-sharing tetrahedra, is a prototypical geometrically frustrated lattice (Fig. 1b). It has attracted much attention as a promising toy model for realizing emergent phenomena due to its rich phenomenology of correlated and topological phases, coupled with its unique underlying lattice geometry \cite{guo_three-dimensional_2009,yang_topological_2010,zhou_weyl_2019,jiang_giant_2021}. The intimate interplay between strong SOC ($\lambda$) and electron correlation ($U$) in the pyrochlore lattice engenders a great diversity of correlated topological phases such as multipolar ordering and spin liquids in the strongly interacting limit, as well as the axion insulator and topological Mott insulator phases in the moderately interacting limit (Fig. 1a) \cite{pesin_mott_2010,witczak-krempa_correlated_2014}. Interestingly, even in the weakly interacting limit, a subtle balance between competing energy scales including bandwidth, spin-orbit coupling, and crystal fields can induce various topological electronic band features such as the 3D Dirac dispersion and the quadratic band crossing, giving rise to the 3D Dirac semimetal and Luttinger-Abrikosov-Beneslavskii semimetal phases, respectively (Supplementary SFig. 1) \cite{yang_topological_2010,moon_non-fermi-liquid_2013,jiang_giant_2021}. Recently, 3D flat bands originating from 3D destructive interference between electron wave functions have been explored in several pyrochlore materials, but these studies have focused on the weak SOC limit to minimize their bandwidth and realize the ideal 3D flat bands. \cite{huang_three-dimensional_2023,wakefield_three-dimensional_2023}. Despite these earlier efforts, a complete characterization of their topological nature is still lacking due to their inherently weak SOC of previously studied materials. Establishing the topological band structure of the strongly spin-orbit coupled pyrochlore lattice in the itinerant limit is thus a timely and exciting challenge of quantum matter research for realizing diverse correlated topological phenomena in the geometrically frustrated lattice systems. 

Here, we focus on the normal state of the pyrochlore superconductor RbBi$_{\textrm{2}}$ (space group Fd$\bar{3}$m, No. 227) to closely and directly examine the band topology of the pyrochlore lattice \cite{philip_bismuth_2021,gutowska_superconductivity_2023}. RbBi$_{\textrm{2}}$ is composed of a combination of Bi pyrochlore and Rb diamond networks (see Fig. 1b). Thanks to the heavy Bi element, RbBi$_{\textrm{2}}$ is a promising material candidate for exploring the topological electronic band structure of pyrochlore lattice in the strong SOC limit (Fig. 1a)  \cite{pesin_mott_2010,moon_non-fermi-liquid_2013,witczak-krempa_correlated_2014,philip_exploring_2023}. Furthermore, it deserves much attention since the potential coupling between topological states and bulk superconductivity ($T_{\textrm{c}}$ $\sim$ 4 K) may lead to exotic topological superconductivity \cite{kobayashi_topological_2015,flensberg_engineered_2021}.

From a symmetry standpoint, the Bi pyrochlore network of RbBi$_{\textrm{2}}$ is described by the $T_{d}$ point group of tetrahedra with twofold rotation ($\textit{C}$$_{2}$) and fourfold screw rotation symmetries ($\textit{S}$$_{4}$) along [001], threefold rotation ($\textit{C}$$_{3}$) symmetry along [111], mirror symmetry ($\sigma$$_{d}$) with respect to the (110) plane, spatial inversion ($\textit{P}$), and time-reversal symmetry ($\textit{T}$).
The complex symmetries of RbBi$\tw$ allow various nodal structures to appear.
We fully characterize the symmetry-allowed nodal structures in the paramagnetic space group (SG) 227 with and without SOC, and summarize the results in Fig. 1c.
In the absence of SOC, a 4-fold degeneracy is always protected at the high symmetry points $X$ and $W$, and along the high symmetry line connecting these points.
On the other hand, depending on the irreducible representation (irrep), the $\Gamma$ point can host 2-, 4-, or 6-fold degeneracies, and the L point can be either 2- or 4-fold degenerate.
Similarly, the high symmetry lines $\Gamma$-X and $\Gamma$-L can be either 2-fold or 4-fold degenerate.
Moreover, an accidental degeneracy may induce 4-fold degenerate nodal points on the high symmetry planes $\Gamma$-X-W-K and $\Gamma$-L-K, or 4-fold degenerate nodal lines at a general momentum (in the degeneracy counting, the spin degrees of freedom are always considered).

Even in the presence of strong SOC, the rich symmetries of this system enable other types of nodal degeneracies (Fig. 1c).
Since $\textit{P}$ and $\textit{T}$ symmetries exist simultaneously, the electronic state always has a two-fold spin degeneracy at every momentum.
Furthermore, the nonsymmorphic $\textit{S}$$_{4}$ symmetry with fourfold rotation and 1/4 fractional translation (Fig. 1d) is associated with symmetry-enforced 3D Dirac band crossings at the high symmetry X point on the Brillouin zone boundary, a scenario first proposed in the $\beta$-cristobalite BiO$_{\textrm{2}}$, which belongs to the same space group as RbBi$_{\textrm{2}}$, but not yet demonstrated \cite{young_dirac_2012} (see also Supplementary SFig. 2).
The relativistic electronic structure of RbBi$_{\textrm{2}}$ calculated by density functional theory (DFT) and shown in Fig. 1f (see also Supplementary SFig. 3) captures these multiple fourfold band crossing points originating from Bi p orbitals, as indicated by red arrows.
These Dirac points are expressed in the four-dimensional $\bar{X}_{5}$ irreducible representation. The symmetry-enforced Dirac fermions of RbBi$_{\textrm{2}}$ exhibit large energy scales that are experimentally detectable due to the strong SOC \cite{huang_three-dimensional_2023,wakefield_three-dimensional_2023}. 
In addition, the Dirac band crossing marked by a blue arrow at the non-high symmetry point on the $\Gamma$-L line (the $\textit{C}$$_{3}$ rotation axis in the momentum space) is driven by an accidental band crossing between the $\bar{\Lambda}_{45}$ (green solid line) and $\bar{\Lambda}_{6}$ band (orange solid line) accompanied by band inversion in the presence of SOC, similar to the case of Na$_{\textrm{3}}$Bi and Cd$_{\textrm{3}}$As$_{\textrm{2}}$ \cite{liu_discovery_2014,liu_stable_2014,xu_observation_2015}.
In contrast to the symmetry-enforced band crossing, this degenerate point is allowed by $\textit{C}_{\textrm{3}}$ rotation symmetry (see Fig. 1e and Supplementary SFig. 2). We note that the band inversion is formed by strong SOC, eventually forming a Dirac band crossing (see Supplementary SFig. 3). 
At the $\Gamma$ point, a quadratic band crossing with the four-dimensional $\bar{\Gamma}_{11}$ irreducible representation, one of the representative electronic structures of the pyrochlore lattice protected by cubic and time-reversal symmetry, was identified near $E = -$2 eV. 
To experimentally demonstrate these theoretically expected topological band features of the pyrochlore superconductor RbBi$_{\textrm{2}}$, we performed high energy and momentum resolution angle-resolved photoemission spectroscopy (ARPES) experiments on the RbBi$_{\textrm{2}}$ single crystals, and comprehensively mapped out the electronic band structure across the 3D Brillouin zone.

In Fig. 2, we first present the symmetry-enforced 3D Dirac fermions at the X point. Fig. 2a shows the measured Fermi surface of RbBi$_{\textrm{2}}$ at $k_{\textrm{z}}$ = 0 obtained from its (111) cleaved surface (see Supplementary SFig. 4 for reference to high symmetry points in the Brillouin zone). Multiple hexagonal, triangular, and elliptical Fermi pockets are observed at the Fermi surface. Notably, the high-symmetry ARPES spectrum along the K-X-K momentum direction displays multiple distinct Dirac band crossings at the X point as shown in Fig. 2b. There are three Dirac points at $E$ = $-$0.45, $-$1, $-$1.8 eV, marked with red, blue, and yellow arrows, respectively. These experimental band dispersions are consistent with the DFT band structure calculations once $k_{\textrm{z}}$ broadening effects are included as shown in Fig. 2c (see Supplementary SFig. 5 for the estimation of the $k_{z}$ broadening effect). These symmetry-enforced Dirac dispersions can be seen more apparently in the stacked energy distribution curves (EDCs) (Fig. 2d). The experimentally measured Dirac velocities along the X-K momentum direction were estimated to be $v_{\textrm{X-K}}^{\textrm{DP1}}$ = 3.11 eV$\cdot${\AA} or 4.72 $\times$ 10$^{\textrm{5}}$ m/s, $v_{\textrm{X-K}}^{\textrm{DP2}}$ = 2.1 eV$\cdot${\AA} or 3.19 $\times$ 10$^{\textrm{5}}$ m/s, and $v_{\textrm{X-K}}^{\textrm{DP3}}$ = 1.94 eV$\cdot${\AA} or 2.95 $\times$ 10$^{\textrm{5}}$ m/s, which agree well with the theoretical prediction (see supplementary Table 1). In Fig. 2e-h, we show photon energy-dependent ARPES spectra to visualize the $k_{\textrm{z}}$ momentum dependence of the topmost Dirac point (DP1) (see supplementary SFig. 6 and 7 for the $k_{\textrm{z}}$ momentum dependence of DP2 and DP3). At $k_{\textrm{z}}$ = $-$0.4 $\pi$/c (Fig. 2e), there is a clear energy gap near $-$0.4 eV binding energy. As the $k_{\textrm{z}}$ momentum approaches the X point ($k_{\textrm{z}}$ = 0 {\AA}$^{\textrm{$-$1}}$), the spectral weight fills the gap, eventually showing the gapless band dispersion at the X point as shown in Fig. 2f and 2g. The spectral weight is again gapped as $k_{\textrm{z}}$ is increased past the X point (Fig. 2h). The $k_{\textrm{z}}$ momentum dependence of the DP1 shows an excellent agreement with DFT simulations (Fig. 2i-k). These 3D Dirac points arising at the high symmetry point on the Brillouin zone boundary (Fig. 2l) represent direct evidence of the predicted symmetry-enforced 3D Dirac fermions.

The degeneracy of the symmetry-enforced Dirac points, protected by the inherent symmetries of the pyrochlore lattice, is lifted along the nodal line between the X and W points due to the presence of SOC (see fig. 1c, supplementary SFig. 1, and SFig. 2). As shown in Fig. 3a, the Dirac points DP1 and DP2, captured in the high symmetry ARPES spectrum along the K-X-K momentum direction, are split between the X and W points and they are linked to the gapped dispersion at the W point along the red dashed curves obtained from the DFT calculation. The experimental observation of the nodal line splitting is in agreement with the electronic structure calculation in Fig. 3b except for some spectral weight spillover due to the finite $k_{\textrm{z}}$ broadening effect. As illustrated in Fig. 3c, through a combination of ARPES experiments and DFT, we can effectively interpolate the Dirac dispersion in the K-X-W momentum plane at the Brillouin zone boundary, which is inaccessible with (111) cleaved surface.

As anticipated by the band structure calculations in Fig. 1f, our ARPES measurements explicitly capture the Dirac point induced by band inversion between $\Gamma$ and L points. In Fig. 4, we present the systematic photon energy-dependent ARPES spectra to trace the 3D Dirac band crossing at the non-high symmetry point on the $k_{\textrm{z}}$ momentum axis (see supplementary Fig. S8). The first (Fig. 4a-d), second (Fig. 4e-h), and third row (Fig. 4i-l) show the ARPES spectra, corresponding curvature plots, and the calculated electronic band structure, respectively. As shown in Fig. 4a, a quadratic band crossing at the $\Gamma$ point ($k_{\textrm{z}}$ = 0), which is one of the archetypal band topological features of the pyrochlore lattice, was observed near the $-$2 eV binding energy. The curvature plots and calculated band structure presented in Fig. 4e and 4i are remarkably consistent. This fourfold degenerate point splits away from the $\Gamma$ point. As shown in Fig. 4b and 4f, the upper branch, separated from the quadratic band crossing, displays a hole-like dispersion with a quadratic momentum dependence (red dashed curve) at $k_{\textrm{z}}$ = 0.92$\pi$/c. Although this hole-like band is located in the vicinity of another band that lies near $-$0.6 eV binding energy, a small energy gap between them was identified in the curvature plot consistent with DFT calculation (see Fig. 4f and 4j). While the ARPES spectrum at $k_{\textrm{z}}$ = 0.96$\pi$/c (Fig. 4c) exhibits very similar spectral weight to that at $k_{\textrm{z}}$ = 0.92$\pi$/c (Fig. 4b), its curvature plot in Fig. 4g provides clear evidence for the closing of the energy gap between those two bands, resulting in a gapless linear band crossing (red dashed line) as predicted by theoretical calculations (Fig. 4k). The in-plane Dirac velocity of this accidental Dirac point was estimated to be $v_{\textrm{x}}$ = 1.93 eV$\cdot${\AA} or 2.93 $\times$ 10$^{\textrm{5}}$ m/s consistent with DFT (see supplementary Table 1). Lastly, we observed that the linear crossing transforms into a gapped dispersion at the L point ($k_{\textrm{z}}$ = $\pi$/c) on the Brillouin zone boundary (see Fig. 4d, 4h, and 4l). This systematic $k_{\textrm{z}}$ momentum evolution, with the gap closing at a non-high symmetry point allowed by rotation ($\textit{C}_{n}$), spatial inversion ($\textit{P}$), and time-reversal ($\textit{T}$) symmetries, embodies the typical tendency of the accidental band crossing associated with band inversion.

This thorough symmetry analysis and comprehensive ARPES study of the three-dimensional electronic band structure of RbBi$_{\textrm{2}}$ provide direct and robust theoretical and experimental proof for symmetry-enforced 3D Dirac fermions, quadratic band crossing, and coexisting accidental Dirac band crossing thanks to the strong SOC of Bi pyrochlore network. The symmetry-enforced 3D Dirac fermions observed in this work differ in fundamental ways from the accidental Dirac band crossings which have been experimentally verified in Na$_{\textrm{3}}$Bi, Cd$_{\textrm{3}}$As$_{\textrm{2}}$, and others. In the presence of $\textit{P}$ and $\textit{T}$ symmetries, nonsymmorphic symmetry imposes the fourfold degenerate points at the high symmetry points on the Brillouin zone boundary. Therefore, the symmetry-enforced 3D Dirac points are robust in crystals belonging to the nonsymmorphic space group and cannot be lifted unless the nonsymmorphic symmetry is broken. In contrast, an accidental Dirac band crossing protected by $\textit{C}_{n}$ (n = 3,4,6) rotational symmetry is not always guaranteed even if the crystals are $\textit{C}_{n}$ symmetric \cite{wang_dirac_2012,yang_classification_2014}. For example, if the energy scale of SOC is not enough to generate the band inversion, an accidental Dirac band crossing cannot appear even in the $\textit{C}_{n}$ symmetric crystals. Therefore, the band inversion-induced 3D Dirac fermion in RbBi$_{\textrm{2}}$ is crucially enabled by the strong SOC. In the case of the quadratic band crossing, our measurements clearly show both the upper and lower bands touching with quadratic dispersion at the $\Gamma$ point. This can be contrasted to the previous ARPES study of pyrochlore iridate Pr$_{\textrm{2}}$Ir$_{\textrm{2}}$O$_{\textrm{7}}$ in which only the lower degenerate band below the Fermi energy was unambiguously identified \cite{kondo_quadratic_2015}. 
It is worth noting that various topological phases can be evoked by applying external perturbations due to the presence of these topological nodal structures. For instance, the fourfold degeneracy of the symmetry-enforced Dirac point can be lifted into several twofold Weyl points by breaking either inversion or time-reversal symmetries while maintaining a gapless dispersion \cite{young_dirac_2012}. In addition, if we eliminate the nonsymmorphic symmetry by applying external stimuli such as strain or external fields, the symmetry-enforced Dirac fermions will gain a finite mass and lead to topologically trivial or non-trivial insulating phases \cite{yang_topological_2010,young_dirac_2012,philip_exploring_2023}. Further, a diversity of correlated topological states can emerge by appropriate material design through chemical substitution or doping to engineer the energy scales of SOC and electron correlations. Using such external perturbations to manipulate the underlying symmetries and tune the balance of competing energy scales, one can create a variety of novel optical, electronic, and magnetic properties of such quantum topological states in the pyrochlore lattice, including chiral charge pumping \cite{xiong_evidence_2015}, axion electrodynamics \cite{gao_layer_2021-1,ahn_theory_2022,qiu_axion_2023-1}, and fractionalized quantum spin Hall effect \cite{young_fractionalized_2008,pesin_mott_2010,witczak-krempa_correlated_2014} in the magnetic Weyl semimetal, axion insulator, and topological Mott insulator phases, respectively. Moreover, the superconducting order in RbBi$_{\textrm{2}}$ makes this material a unique platform for studying topological superconductivity. While the theory has predicted the exotic superconducting properties in the Dirac semimetals \cite{kobayashi_topological_2015} and surface superconductivity was reported in the 3D Dirac semimetal Cd$_{\textrm{3}}$As$_{\textrm{2}}$ \cite{wang_observation_2016,aggarwal_unconventional_2016}, their ground states are non-superconducting under ambient conditions \cite{he_pressure-induced_2016}. In contrast, thanks to the intrinsic bulk superconductivity of RbBi$_{\textrm{2}}$ and the prospective intimate coupling with its two different types of Dirac fermions, exclusive topological superconducting properties can arise in this pyrochlore superconductor. Future studies with complementary experimental techniques, such as scanning tunneling microscopy (STM) \cite{wang_evidence_2018} and high-resolution ARPES \cite{zhong_nodeless_2023}, will enable shedding light onto the nature of topological superconductivity in this material. 


\section*{Methods}

\subsection*{Single-crystal growth}
Single crystals of RbBi$_{\textrm{2}}$ were grown using a self-flux method \cite{canfield_growth_1992} similar to described in our recent report \cite{gutowska_superconductivity_2023}, with an excess of Bi (with Rb:Bi ratio varied between 1:9 and 3:17). Rubidium pieces (99.99\% pure, Onyxmet Poland) were scooped into alumina crucibles and covered with Bismuth pieces (99.99\%, Onyxmet). Crucibles were loaded into silica glass tubes along with plugs of quartz wool used for separation of crystals from the flux. All manipulations were performed inside an Argon-filled glovebox, as elemental Rubidium ignites spontaneously in air. 

Tubes were evacuated, back-filled with high-purity Ar, and subsequently sealed. Resulting ampoules were put in a box furnace, heated to 560 $^{\circ}$C at 50 $^{\circ}$C/h, soaked for 8 h, and then slowly cooled to 315 $^{\circ}$C at 2.5 $^{\circ}$C/h. They were then centrifugated to remove the excess flux. Octahedral crystals with sizes up to a few millimeters were obtained. Powder x-ray diffraction confirmed their composition and crystal structure. A small amount of Bi was observed in the powder patterns, which comes both from the leftover Bi flux and from the decomposition of the RbBi$_{\textrm{2}}$ phase in moist air \cite{winiarski_superconductivity_2016,gutowska_superconductivity_2023}.

\subsection*{Angle-resolved photoemission spectroscopy (ARPES)}
ARPES experiments were performed at Beamline 7.0.2 (MAESTRO) of the Advanced Light Source (ALS). RbBi$_{2}$ single crystals were taken out from the sealed quartz ampoules and mounted on the sample holders inside the glovebox to avoid degradation. The mounted RbBi$_{2}$ single crystals were transferred directly from glovebox to load-lock chamber through \textit{in-situ} transfer path. RbBi$_{2}$ single crystals were cleaved inside an ultra-high-vacuum (UHV) ARPES chamber ($\sim$4 $\times$ 10$^{-11}$ torr). The photoelectrons were collected by a hemispherical electron analyzer equipped with a photoelectron deflector. ARPES data were obtained from (111) cleaved surface with linear horizontal polarization (p-polarization). The energy and momentum resolutions were approximately 20 meV and 0.01 \AA$^{-1}$, respectively. To determine the high symmetry points along the $k_{\textrm{z}}$ momentum, photon energy-dependent ARPES measurements were performed from 70 eV to 175 eV. Inner potential $V_{\textrm{0}}$ = 10 eV was used to estimate the $k_{\textrm{z}}$ momentum.

\subsection*{Density functional theory calculations}
To perform the density functional theory (DFT) calculations, we used the code implemented in Vienna Ab initio Simulation Package (VASP) ~\cite{vasp1,vasp2} to derive the electronic states and band structures. 
The electronic ground state is converged within the Projector augmented wave method (PAW) \cite{PAW} pseudopotential formalism, exchange-correlation energies parametrized by Perdew, Burke and Ernzerhof (PBE)~\cite{pbe}, a 520 eV cutoff energy for the plane-wave-basis set, and a $6 \times 6 \times 6$  Monkhorst-Pack grid sampling~\cite{MP_grid} in the reciprocal space.
Relativistic spin-orbit couplings were included in obtaining the electronic band structure.
To efficiently capture the bands in the full Brillouin zone, we computed the bands on a finer $91 \times 91 \times 91$ momentum grid used for deriving the interpolation and smearing of the electronic structure.

\subsection*{Acknowledgements}
This work was supported by the Air Force Office of Scientific Research (AFOSR) under grant FA9550-22-1-0432. S.F and J.G.C acknowledge support from the Gordon and Betty Moore Foundation EPiQS Initiative, Grant No. GBMF9070. This research used resources of the Advanced Light Source, which is a DOE Office of Science User Facility under contract no. DE-AC02-05CH11231. The work at Gdansk Tech. was supported by the National Science Centre (Poland; Grant UMO-2018/30/M/ST5/00773). J.K, Y.Q, B.-J.Y. were supported by the Institute for Basic Science in Korea (Grant No. IBS-R009-D1), Samsung Science and Technology Foundation under Project number SSTF-BA2002-06, the National Research Foundation of Korea (NRF) grant funded by the Korean government (MSIT) (No. 2021R1A2C4002773, and No. NRF-2021R1A5A1032996). 

\subsection*{Author contributions}
D.O. and R.C. conceived the experiments. D.O. and M.K. performed the synchrotron based ARPES experiments with support from C.J., A.B., and E.R.; S.F. performed the electronic structure calculation based on density functional theory with help from J.C.; D.O., J.K., Y.Q., B.-J.Y., and L.F. conducted a symmetry analyses. T.K. and M.J.W. synthesized and characterized the single crystals. D.O., J.K., B.-J.Y. and R.C. wrote the manuscript with contributions from all co-authors.

\pagebreak

\begin{figure}[h]
	\includegraphics[width=16.4cm]{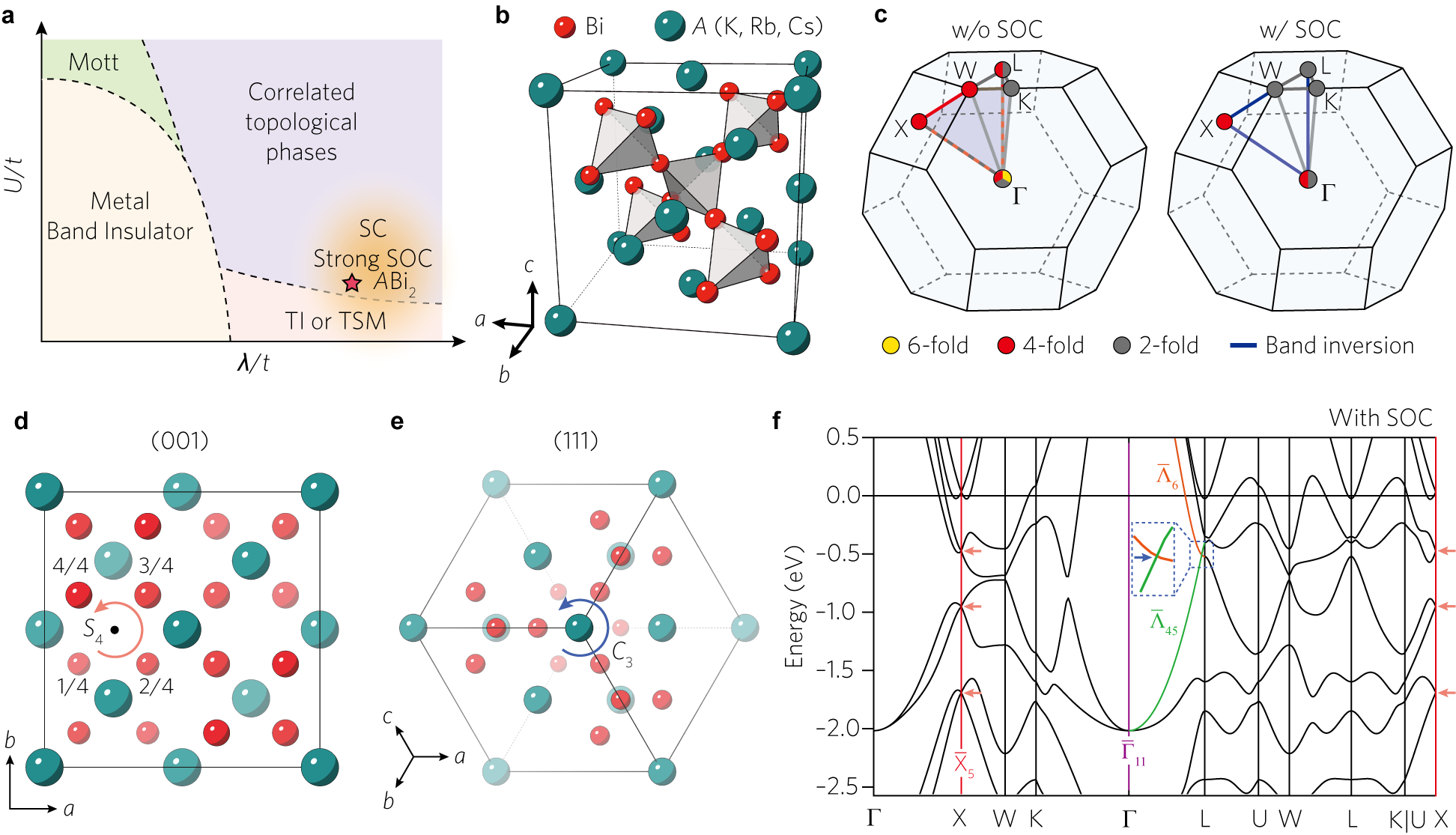}
        \caption{\textbf{ \textit{U} vs $\lambda$ phase diagram, crystal structure, and  electronic structure of RbBi$_{2}$}. 
        \textbf{a,} Phenomenological phase diagram of correlated topological materials as a function of electron correlation \textit{U} and spin-orbit coupling $\lambda$. RbBi$_{\textrm{2}}$ is expected to be located in the orange-shaded region due to its strong spin-orbit coupling. TI, TSM, and SC are abbreviations for topological insulators, topological semimetals, and superconductors, respectively.
        \textbf{b,} Crystal structure of RbBi$_{\textrm{2}}$. Green and red spheres correspond to the Rb and Bi atoms, respectively. Rb and Bi atoms form a diamond and pyrochlore sublattice, respectively. Polyhedrons composed of Bi atoms show the corner-sharing tetrahedrons of the pyrochlore lattice. 
        \textbf{c,} Bulk Brillouin zone and nodal structure of SG 227. Yellow, red, and gray filled circles and lines indicate 6-, 4-, and 2-fold degeneracy including spin degeneracy enforced by symmetry. The blue lines and shared regions indicate a possible 4-fold accidental degeneracy originating from band inversion. 
        \textbf{d,e,} Views of crystal structure along (\textbf{d}) [001] and (\textbf{e}) [111] directions. Transparency reflects the height of the atoms. Red and blue rotating arrows indicate the fourfold screw rotation (\textit{S}$_{\textrm{4}}$) and threefold rotation (\textit{C}$_{\textrm{3}}$) symmetries, which protect the symmetry-enforced (essential) and accidental band crossing, respectively. 
        \textbf{f,} Theoretical electronic structure of RbBi$_{\textrm{2}}$ calculated by density functional theory (DFT). Symmetry-enforced Dirac band crossings at the X point and accidental Dirac band crossing near the L point are marked by red and blue arrows, respectively.}
\end{figure}

\begin{figure}[h]
	\includegraphics[width=16.53cm]{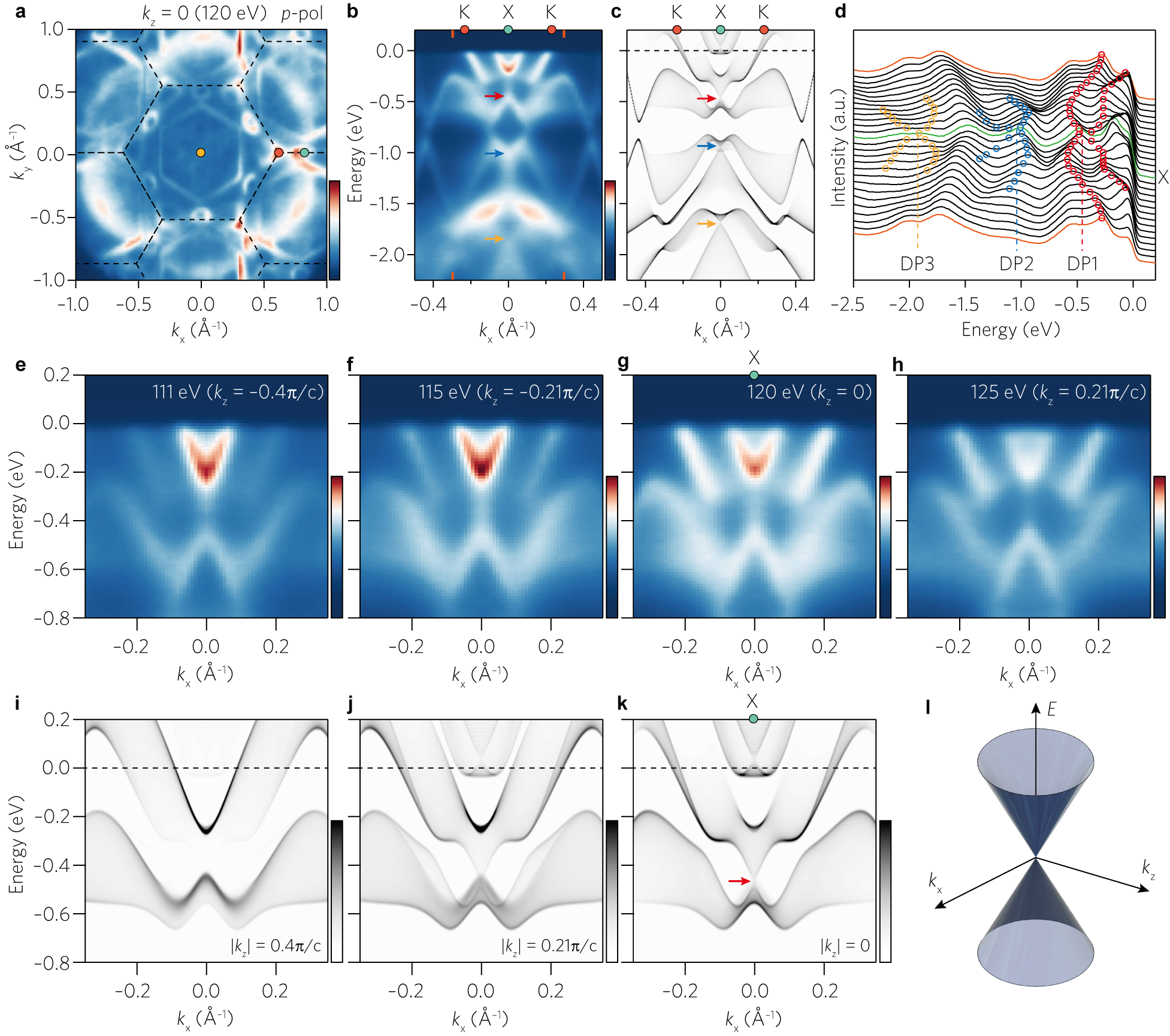}
	\caption{\textbf{Symmetry-enforced Dirac fermions at X point.}
        \textbf{a,} Fermi surface of RbBi$_{\textrm{2}}$ at $k_{\textrm{z}}$=0. Black dashed lines show the Brillouin zone boundaries. Yellow, red, and green circles correspond to the $\Gamma$, K, and X points, respectively.
        \textbf{b,c,} Experimental and simulated ARPES spectrum along the K-X-K direction. The measured ARPES spectrum was symmetrized with respect to the X point. A finite $k_{\textrm{z}}$ broadening effect was taken into account for the simulated spectrum. Red, blue, and yellow arrows indicate the Dirac points 1 (DP1), 2 (DP2), and 3 (DP3), respectively. 
        \textbf{d,} Stacked energy distribution curves along the K-X-K direction. Red, blue, and yellow empty circles and dashed lines mark the Dirac dispersion and Dirac points, respectively.
        \textbf{e-h,} Photon energy-dependent ARPES spectra near DP1 showing 3D Dirac dispersion. 
        \textbf{i-k,} Simulated electronic structures at the different $k_{\textrm{z}}$.
        \textbf{l,} Schematic of the projected symmetry-enforced Dirac band crossing into the $k_{\textrm{x}}$-$k_{\textrm{z}}$-$E$ space.
        Photon energy-dependent ARPES spectra for DP2 and DP3 is shown in supplementary information Fig. S6.}
\end{figure}

\begin{figure}[h]
	\includegraphics[width=15.65cm]{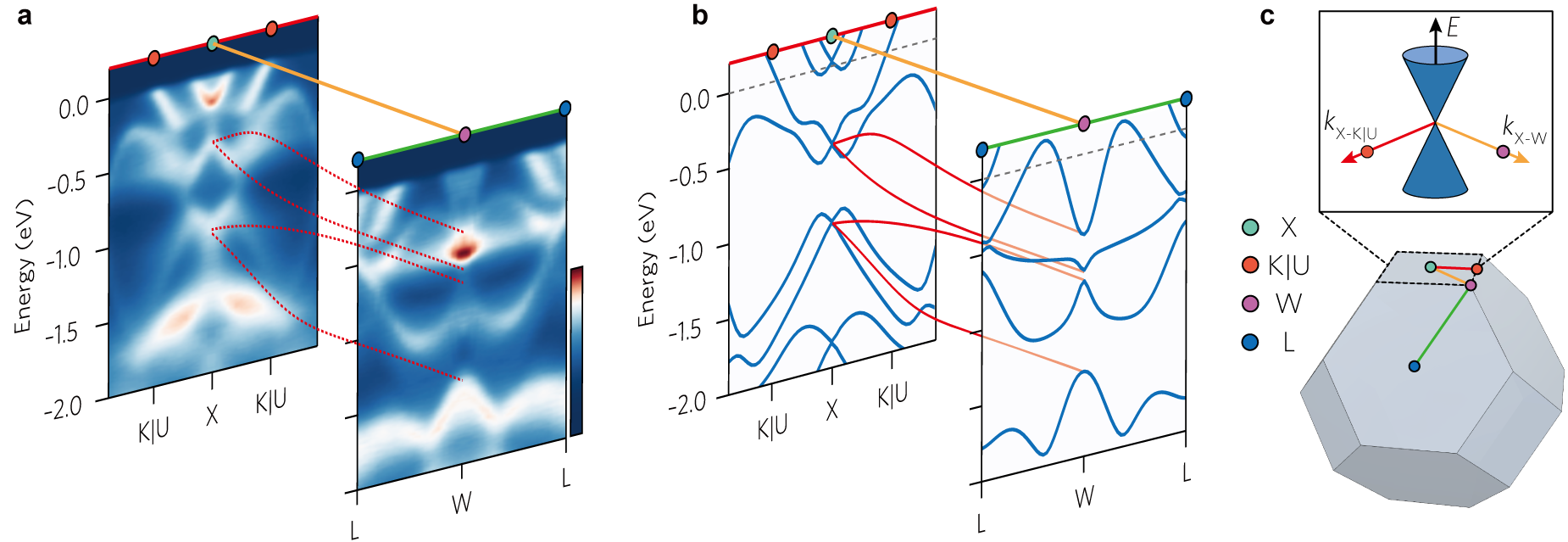}
        \caption{\textbf{Lifted nodal line degeneracy between X and W points.}
        \textbf{a,b,} ARPES spectra (\textbf{a}) and DFT band structure (\textbf{b}) along K|U-X-K|U (left) and L-W-L (right) momentum directions. The measured spectra were symmetrized with respect to the X and W points. Red dashed and solid curves in (\textbf{a}) and (\textbf{b}) represent the calculated nodal line splitting between the X and W points, which cannot be obtained with the (111) cleaved surface. Yellow, green, and red solid straight lines exhibit the high symmetry momentum direction between X-W, L-W, and X-K|U points, respectively.
        \textbf{c,} Schematic of the projected symmetry-enforced Dirac band crossing centered at the X point into the $k_{\textrm{x}}$-$k_{\textrm{y}}$-$E$ space. Green, red, purple, and blue circles correspond to the X, K|U, W, and L points, respectively. 
        }
\end{figure}

\begin{figure}[h]
	\includegraphics[width=12.9cm]{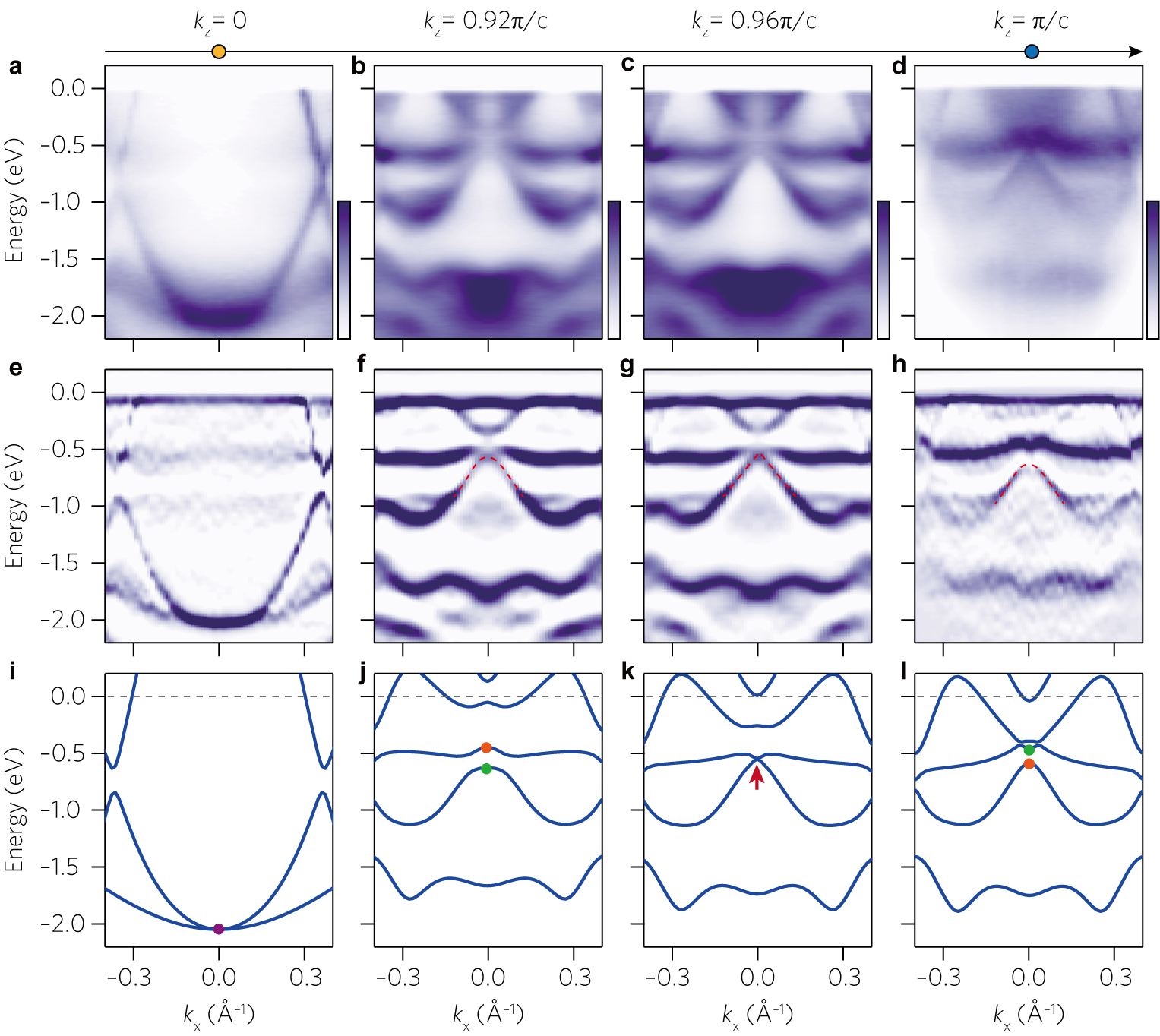}
    \caption{\textbf{Quadratic band crossing and band inversion induced Dirac fermion.}
    \textbf{a-d,} Photon energy-dependent ARPES spectra from (\textbf{a}) $\Gamma$ point ($k_{\textrm{z}}$=0) (yellow circle) to (\textbf{d}) L point ($k_{\textrm{z}}$=$\pi$/c) (blue circle). \textbf{a,d} and \textbf{b,c} were obtained from the center of the second and first Brillouin zone, respectively. The momentum positions from which these spectra were extracted in the three-dimensional Brillouin zone are shown in supplementary information Fig. S8. \textbf{b} and \textbf{c} were symmetrized with respect to $k_{\textrm{x}}$=0.
    \textbf{e-h,} Corresponding curvature plots for ARPES spectra in (\textbf{a-d}).
    \textbf{i-l,} Calculated electronic band structure of RbBi$\textrm{2}$ at $k_{\textrm{z}}$=0, 0.92$\pi$/c, 0.96$\pi$/c, and $\pi$/c, respectively. The purple circle marks the quadratic band crossing point with the four-dimensional $\bar{\Gamma}_{\textrm{11}}$ irreducible representation. Green and orange circles represent two-dimensional $\bar{\Lambda}_{\textrm{45}}$ and $\bar{\Lambda}_{\textrm{6}}$ irreducible representations, respectively. The red arrow indicates the band inversion induced Dirac point protected by three-fold rotation symmetry.}
\end{figure}

\clearpage

\end{document}